\documentclass[twocolumn,showpacs,preprintnumbers,amsmath,amssymb]{revtex4-1}
\usepackage{graphicx}
\usepackage{dcolumn}
\usepackage{bm}
\usepackage{epstopdf}
\usepackage{mhchem}
\usepackage{color}
\begin{document}

\title{Hydration-induced local diffusivity heterogeneity modulates aqueous \ce{H2 + OH} reaction kinetics}

\author{Ming-Gen Li$^{1,}$}
\email{mgli@stu.edu.cn}
\author{Shi-Yuan Zhang$^1$}
\author{Alessio Zaccone$^2,$}
\email{alessio.zaccone@unimi.it}

\affiliation{$^1$ Department of Physics, College of Science, Shantou University, Shantou, Guangdong 515063, China
\\$^2$ University of Milan, Department of Physics, via Celoria 16, 20133, Milano, Italy}

\date{\today}

\begin{abstract}
At room temperature, a Kramers reaction rate calculated with homogeneous solvent friction exceeds the measured rate of the aqueous reaction \ce{H2 + ^.OH -> H^. + H2O} by about an order of magnitude.
We show that spatially correlated hydration-induced diffusivity fluctuations increase effective friction and suppress the predicted escape rate.
Estimates based on hydration-shell properties bring the theoretical rate closer to experiment.
The suppression weakens with increasing temperature, linking local solvent heterogeneity to activated aqueous reaction kinetics.
\end{abstract}
\maketitle

The hydroxyl radical (\ce{^.OH}) plays important roles in biological processes \cite{dizda,suix}, environmental remediation \cite{xujk}, and therapeutic applications \cite{tang}, with hydrogen abstraction among its key reaction pathways.
The prototypical four-atom reaction
\ce{H2 + ^.OH\rightarrow H^. + H2O} provides a benchmark for elementary reaction dynamics \cite{smith,zhang}.
Its rate constants have been measured over a broad temperature range by pulse radiolysis \cite{marin}.
In solution, conventional Kramers theory describes transition-state barrier crossing along a reaction
coordinate under homogeneous solvent friction \cite{krame,hangg,garci,polla2}.
At room temperature ($298~\mathrm{K}$), however, the Kramers rate estimate for this aqueous reaction, $\sim 4.9\times10^{8}\,
{\rm L\,mol^{-1}\,s^{-1}}$ \cite{rouss}, exceeds the experimental value of $\sim 4.2\times10^{7}\,
{\rm L\,mol^{-1}\,s^{-1}}$ \cite{buxto,siwek}, by more than an order of magnitude.
This discrepancy suggests that the homogeneous-friction description may omit important solvent contributions \cite{bagch2}.

Previous studies have shown that \ce{^.OH} is strongly coupled to its local hydration environment, where
the surrounding water molecules form a finite, dynamically fluctuating solvation cage \cite{brana,kjell,bagch1}.
Cage fluctuations and rearrangements can produce heterogeneous dissipative coupling between the reactive
species and the solvent \cite{codor,wojci,hadiz,azacc}.
Such heterogeneity is absent from conventional homogeneous-friction Kramers treatments \cite{addad1,addad2,reacc,pgbha}.
The central question is whether hydration-induced dissipation heterogeneity, characterized by properties of the \ce{^.OH} hydration environment, can quantitatively account for the room-temperature rate discrepancy.

In this Letter, we develop a coarse-grained kinetic model for the reaction \ce{H2 + ^.OH\rightarrow H^. + H2O} by representing hydration-induced dissipation heterogeneity as spatially correlated diffusivity fluctuations.
The fluctuation strength is estimated from water mobility statistics at the molecular scale specified by the hydroxyl-radical hydration number, and the correlation length from hydration-cage geometry.
We derive the resulting renormalized barrier-crossing friction and its effect on the reaction rate.
We find that the friction enhancement substantially reduces the room-temperature discrepancy between the homogeneous-friction prediction and experiment.
The model also predicts that the relative friction
enhancement and rate suppression weaken with increasing
temperature over $298$--$360~\mathrm{K}$.

{\it Model of reaction dynamics with hydration-induced diffusivity heterogeneity}.---Hydration-cage
rearrangements may produce spatially varying dissipation and temporal memory.
Reported \ce{^.OH-H2O} hydrogen-bond lifetimes near 310 K
are $\sim0.03~\mathrm{ps}$ (continuous) and $\sim3.6$--$3.8~\mathrm{ps}$
(intermittent) \cite{pabis}.
A Grote--Hynes interpretation of the ratio of the experimental rate to the
transition-state-theory rate yields an effective unstable-mode time of
$\sim0.40~\mathrm{ps}$ at $298~\mathrm{K}$  \cite{grote,eploa}.
These order-of-magnitude estimates, obtained at nearby temperatures,
motivate distinguishing rapid local fluctuations from
more persistent environmental structure.
We adopt a conditional Markovian approximation:
for a given hydration configuration, the remaining
fast solvent degrees of freedom are treated as
a Markovian thermal bath.
The associated spatial dissipation profile is assumed to remain
approximately static over each transition path.
The system-bath representation and temporal coarse-graining are presented
in the Supplemental Material.
Within this framework, the dynamics along $Q$ are
governed by the Langevin equation \cite{wolfs1,wolfs2,mgli1,azacco,chers},
\begin{equation}
\label{eqlan}
\mu\ddot Q(t) =-\frac{\partial U(Q(t))}{\partial Q}
-\mu\gamma(Q)\dot Q+\sqrt{2\mu k_{\rm B}T\gamma(Q)}\xi(t),
\end{equation}
where $\mu$ is the reduced mass,
$k_B$ is the Boltzmann constant, and $T$ is the temperature.
The position-dependent friction $\gamma(Q)$ characterizes the dissipative coupling
between the reaction coordinate and the aqueous environment.
The noise $\xi(t)$ has zero mean and satisfies
$\langle\xi(t)\xi(t')\rangle_n=\delta(t-t')$ \cite{kubor}, where $\langle\cdot\rangle_n$ denotes
an average over realizations of the noise.
For constant friction $\gamma(Q)=\gamma_0$,
Eq.~(\ref{eqlan}) reduces to the Langevin description
underlying conventional Kramers theory with homogeneous
friction \cite{bruni,eolla}.

\begin{figure}[t]
\centering
\includegraphics[width=0.46\textwidth]{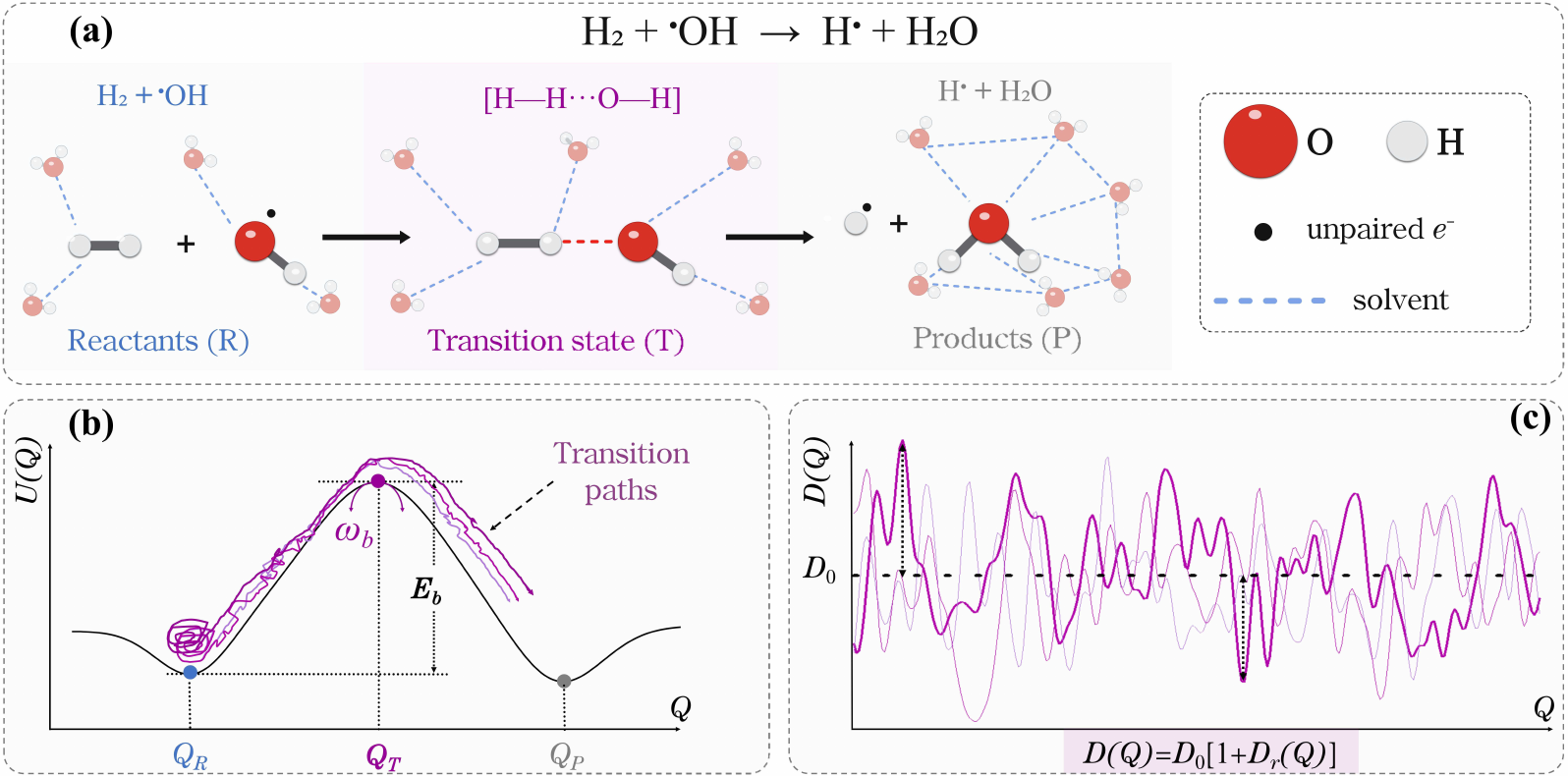}
\caption{\label{model}
Schematic of barrier-crossing dynamics for the aqueous \ce{H2 + ^.OH\rightarrow H^. + H2O}
reaction with hydration-induced diffusivity
heterogeneity.
(a) Reactant, transition-state, and product configurations.
(b) The reaction is described as activated barrier crossing
along the reaction coordinate $Q$ in the potential of mean
force $U(Q)$, with barrier height $E_b$ and unstable-mode
frequency magnitude $\omega_b$.
Transition paths connect the reactant minimum at $Q_R$
to the product region around $Q_P$ through the transition
state at $Q_T$.
(c) Hydration-induced variations in the local diffusivity
$D(Q)$ influence barrier-crossing dynamics along transition
paths.
The dashed line denotes the homogeneous reference diffusivity
$D_0$, and the fluctuations $D_r(Q)$ are specified by Eq.~(\ref{eqcor}).}
\end{figure}

In Fig.~\ref{model}, we illustrate the model for the reaction \ce{H2 + ^.OH\rightarrow H^. + H2O}.
The reaction is projected onto a one-dimensional coordinate $Q$ with
a potential of mean force $U(Q)$, featuring a reactant minimum at $Q_R$, a transition-state barrier
at $Q_T$, and a product region around $Q_P$. The barrier height and unstable-mode
frequency magnitude are $E_b$ and $\omega_b$, respectively.
Barrier-crossing trajectories from the reactant to the
product region define transition paths.
We characterize hydration-induced variations in dissipative
coupling through the diffusivity along $Q$,
\begin{equation}
D(Q)=D_0\left[1+D_r(Q)\right],
\label{eqdif}
\end{equation}
where $D(Q)\equiv k_BT/[\mu\gamma(Q)]$, and $D_0\equiv k_{\rm B}T/(\mu\gamma_0)$ is the homogeneous reference diffusivity.
We treat the dimensionless fluctuation $D_r(Q)$ as a spatial
random field and describe the transition-path ensemble by
averaging the dynamics over realizations of this field.
The field satisfies $\langle D_r(Q)\rangle=0$, with
$D_r(Q)>-1$ ensuring $D(Q)>0$.
Here, $\langle\cdot\rangle$ denotes the average over field realizations.

The finite size of the \ce{^.OH} hydration environment implies that nearby configurations along the reaction
coordinate, at $Q$ and $Q'$, can sample overlapping solvent degrees of freedom.
These configurations are expected to experience correlated fluctuations in dissipative coupling.
We characterize this correlation by $C(|Q-Q'|)\equiv\langle D_r(Q)D_r(Q')\rangle$.
As $|Q-Q'|$ increases, the solvent degrees of freedom sampled at $Q$ and $Q'$ overlap less, leading to weaker correlations between the corresponding diffusivity fluctuations.
As a minimal representation of these short-ranged correlations \cite{sokol}, we adopt
\begin{equation}
\label{eqcor}
C(|Q-Q'|)=g_0\exp\left[-\frac{(Q-Q')^2}{2\lambda^2}\right].
\end{equation}
Here, $g_0=C(0)=\langle D_r^2(Q)\rangle$ quantifies the
fluctuation strength, and $\lambda$ is the characteristic
correlation length along $Q$.
Together, these parameters connect the diffusivity heterogeneity to the local hydration environment.
We next construct coarse-grained estimates of $g_0$ from water mobility statistics at the molecular scale specified by the \ce{^.OH} hydration number and of $\lambda$ from hydration-cage geometry.

\begin{figure}[t]
\centering
\includegraphics[width=0.42\textwidth]{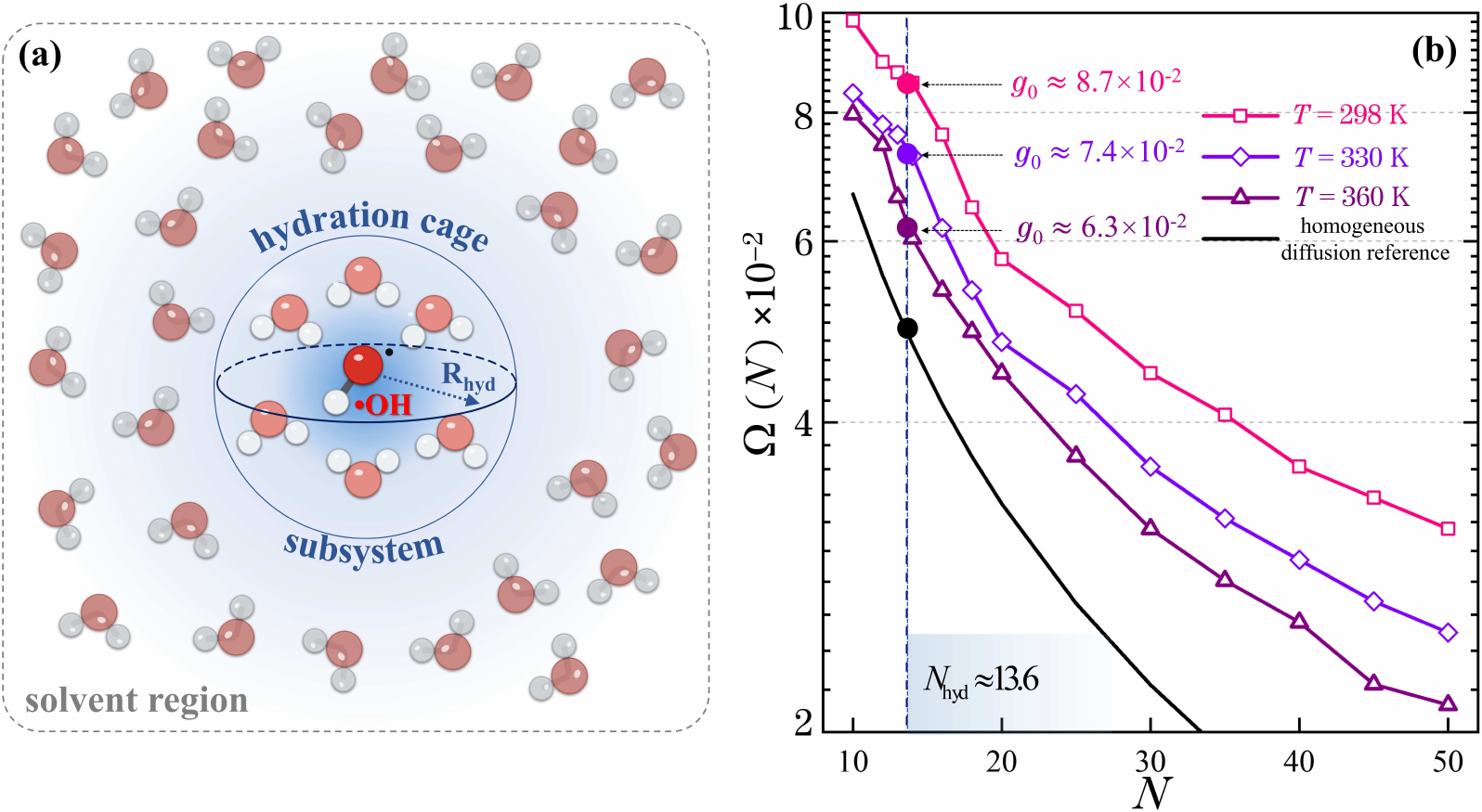}
\caption{\label{hydra}
Estimation of the fluctuation strength $g_0$ and correlation
length $\lambda$ from the \ce{^.OH} hydration environment.
(a) Schematic of the hydration environment, where
$N_{\rm hyd}$ denotes the mean number of water molecules
in the hydration cage and $R_{\rm hyd}$ its characteristic
radius.
(b) Dependence of the local mobility fluctuation measure
$\Omega(N)$ on subsystem size $N$ at several temperatures.
The data are obtained numerically using Eq.~(\ref{eqtim})
with $M_{\rm S}=10$ and $\tau_{\rm obs}=300$ ps.
Simulation details are provided in the Supplemental Material.
The fluctuation strength $g_0$ is estimated by evaluating
$\Omega(N)$ at $N=N_{\rm hyd}$ using Eq.~(\ref{eqg0}).
The solid black line shows the normalized local-MSD variance
$\Omega_{\rm hom}(N)=2/(3N)$ for $N$ independent Brownian
particles with a homogeneous diffusivity in three dimensions.
The correlation length $\lambda$ is estimated from
the hydration-cage geometry using Eq.~(\ref{eqlam}).}
\end{figure}

{\it Hydration-based estimation of $g_0$
and $\lambda$}.---Figure~\ref{hydra}(a) illustrates the local hydration environment of \ce{^.OH}. The surrounding water molecules form a transient, spatially irregular hydration cage, characterized by the mean hydration number $N_{\rm hyd}$ and the characteristic cage radius $R_{\rm hyd}$.

(i) Fluctuation strength $g_0$. We estimate $g_0$ from the size dependence of local mobility fluctuations in water.
For a subsystem containing $N$ water molecules, we
define the local mean-square displacement
between times $t_1$ and $t_2$ as  $\Lambda_S(t_1,t_2)\equiv\sum_{i=1}^{N}
\left|\emph{\textbf{r}}_i(t_1)-\emph{\textbf{r}}_i(t_2)
\right|^2/N$, where $\emph{\textbf{r}}_i(t)$ denotes the position of the $i$th water molecule in the subsystem \cite{scior}.
Its normalized squared
deviation from the system-averaged dynamics is
\begin{equation}
\label{eqome}
\Omega_S(t_1,t_2)=\frac{\left[\Lambda_S(t_1,t_2)-\mathrm{MSD}
(\Delta t)
\right]^2}{\left[\mathrm{MSD}(\Delta t)\right]^2},
\end{equation}
where $\Delta t=|t_1-t_2|$ and $\mathrm{MSD}(\Delta t)$ is the
mean-square displacement averaged over all water molecules in the solvent region.
To characterize the dependence on subsystem size,
we average $\Omega_S$ over time pairs within an observation window $\tau_{\rm obs}$ and
over $M_{\rm S}$ equivalent subsystems, each containing $N$ water molecules:
\begin{equation}
\label{eqtim}
\Omega(N)
=\frac{1}{M_{\rm S}}
\sum_{S=1}^{M_{\rm S}}
\frac{
\int_{\mathcal{A}}dt_1dt_2\,
\Omega_S(t_1,t_2)
}{\int_{\mathcal{A}}dt_1dt_2},
\end{equation}
where the integration domain $\mathcal{A}=
\{(t_1,t_2)|\,0\le t_1<t_2\le \tau_{\rm{obs}}\}$.
For a sufficiently long observation window, $\Omega(N)$ characterizes local
mobility fluctuations at the molecular scale set by $N$.

In the diffusive regime, mean-square displacements over a fixed time interval
scale with the corresponding diffusivity \cite{nikka}.
At the coarse-grained level, we use this scaling
to relate water-subsystem mobility to diffusivity along the
reaction coordinate, adopting
$\Lambda_S(t_1,t_2)\sim D(Q)\Delta t$ and
$\mathrm{MSD}(\Delta t)\sim D_0\Delta t$.
This mapping assumes a common proportionality factor that cancels upon normalization.
Equation~(\ref{eqome}) then becomes
\begin{equation}
\label{norflu}
\Omega_S(t_1,t_2)
\sim\left[\big(D(Q)-D_0\big)/D_0\right]^2=D_r^2(Q).
\end{equation}
To estimate the field variance, we use the average over time pairs
and subsystems in Eq.~(\ref{eqtim}) as a coarse-grained estimate
of the field ensemble average. Selecting the molecular scale
$N=N_{\rm hyd}$ then gives
\begin{equation}
\label{eqg0}
g_0=
\left\langle D_r^2(Q)\right\rangle
\sim
\Omega(N_{\rm hyd}).
\end{equation}
As shown in Fig.~\ref{hydra}(b), numerical simulations show that $\Omega(N)$ decreases with
increasing $N$, reflecting the averaging of local dynamical
fluctuations over larger molecular regions.
For $N$ independent Brownian particles in three dimensions, the normalized local-MSD variance is
$\Omega_{\rm hom}(N)=2/(3N)$.
At $N_{\rm hyd}\approx13.6$, this gives
$4.9\times10^{-2}$, smaller than the coarse-grained estimate
$g_0\approx8.7\times10^{-2}$ at $298~\mathrm{K}$.
Correlated water motion on the hydration-cage scale may
contribute to this excess \cite{scior}.
Repeating this procedure at different
temperatures, with a consistent temporal averaging convention, gives $g_0(T)$.

(ii) Correlation length $\lambda$.---Since local diffusivity is sensitive to the \ce{^.OH} hydration environment, we estimate $\lambda$ using a minimal geometric model of cage renewal \cite{brana,madiz}.
We use the characteristic
length associated with the first loss of an original neighbor
as a coarse-grained proxy for the spatial decorrelation length of the local dissipative environment.
The model idealizes the irregular hydration cage as a spherical region of radius $R$, initially
containing $N$ water molecules with uniformly distributed positions.
The construction therefore characterizes a spatial renewal length.
We use this renewal length as a coarse-grained estimate of the
diffusivity correlation length, which gives
\begin{equation}
\label{eqlam}
\lambda\sim R_{\rm hyd}/N_{\rm hyd},
\end{equation}
where $R_{\rm hyd}$ and $N_{\rm hyd}$ are the hydration-cage radius and hydration number, respectively.
The detailed derivations are given in the Supplemental Material.

{\it Renormalized barrier-crossing friction}.---We now examine how hydration-induced local diffusivity heterogeneity affects barrier crossing near the transition state.
Near the barrier top, $Q\sim Q_T$, the potential $U(Q)$ is
approximated by an inverted parabola, $U(Q)\simeq E_b-\frac{1}{2}\mu\omega_b^2(Q-Q_T)^2$.
In the moderate- to high-friction regime, generally appropriate for reactions in solution, we take the Smoluchowski limit of Eq.~(\ref{eqlan}), eliminating the velocity degree of freedom $\dot Q$.
For a given realization of $D(Q)$, the probability density $P(Q,t)$ obeys \cite{riske,rober,cness}
\begin{eqnarray}
\label{eqsmo}
\hspace{-7mm}\frac{\partial  P}{\partial t}
=\frac{\partial}{\partial Q}\big\{D(Q)\exp[-\beta U(Q)]
\frac{\partial }{\partial Q}\left\{\exp[\beta U(Q)]P\right\}
\big\},
\end{eqnarray}
where $\beta=(k_BT)^{-1}$ and $P$ is the probability density obtained by averaging
over the Langevin noise at fixed diffusivity field.

Substituting $D(Q)=D_0[1+D_r(Q)]$ into Eq.~(\ref{eqsmo}) gives
\begin{eqnarray}
\label{eqsexp}
\begin{aligned}
&\frac{\partial P}{\partial t}
=D_0\frac{\partial^2P}{\partial Q^2}
+\beta D_0
\frac{\partial}{\partial Q}
\left[U'(Q)P
\right] \\&~~
+\beta D_0
\frac{\partial}{\partial Q}
\left[
U'(Q)G(Q,t)
\right]+D_0\frac{\partial}{\partial Q}M(Q,t),
\end{aligned}
\end{eqnarray}
where $U'(Q)\equiv\partial U(Q)/\partial Q$, $G(Q,t)\equiv D_r(Q)P(Q,t)$, and $M(Q,t)\equiv D_r(Q)\partial P(Q,t)/\partial Q$.
Averaging Eq.~(\ref{eqsexp}) over realizations of the diffusivity field yields
\begin{eqnarray}
\label{eqapre}
\begin{aligned}
&\frac{\partial W}{\partial t}
=D_0\frac{\partial^2W}{\partial Q^2}
+\beta D_0
\frac{\partial}{\partial Q}
\left[U'(Q)W
\right] \\
&~~+\beta D_0
\frac{\partial}{\partial Q}
\left[
U'(Q)\langle G(Q,t)\rangle
\right]+D_0\frac{\partial}{\partial Q}\langle M(Q,t)\rangle,
\end{aligned}
\end{eqnarray}
where $W=W(Q,t)\equiv\langle P(Q,t)\rangle$.
Although $\langle D_r(Q)\rangle=0$, the probability density
depends on the diffusivity-field realization.
Consequently, closure approximations are needed for the field--density correlations $\langle G\rangle$ and $\langle M\rangle$.
To evaluate these correlations, we apply the
Furutsu--Novikov identity within a Gaussian statistical
approximation and adopt a local-response closure \cite{klayt}, obtaining
\begin{eqnarray}
\label{sslan}
\langle G(Q,t)\rangle&&\simeq g_0\beta\lambda^2\frac{\partial (U'(Q)W)}{\partial Q}+g_0\lambda^2
\frac{\partial^2W}{\partial Q^2},\\
\label{sslanf}
\hspace{-4mm}\langle M(Q,t)\rangle&&\simeq-g_0\frac{\partial W}{\partial Q}-g_0\beta U'(Q)W+g_0\lambda^2\frac{\partial^3W}{\partial Q^3}\nonumber\\&&~~+g_0\beta\lambda^2\frac{\partial^2 (U'(Q)W)}{\partial Q^2}.
\end{eqnarray}
Detailed derivations of Eqs.~(\ref{sslan}) and (\ref{sslanf}) are provided in the Supplemental Material.

Substituting Eqs.~(\ref{sslan}) and (\ref{sslanf}) into Eq.~(\ref{eqapre}) yields
\begin{eqnarray}
\label{eqafull}
\hspace{-7mm}\frac{\partial W}{\partial t}
\simeq&&
D_0(1-\Xi)\left\{\frac{\partial^2W}{\partial Q^2}
+\beta\frac{\partial}{\partial Q}[U'(Q)W]\right\}
+\Theta[W],
\end{eqnarray}
where, under the parabolic barrier approximation,
\begin{eqnarray}
\label{define}
\hspace{-6mm}\Xi\equiv g_0\left[1-\beta\lambda^2\partial^2 U(Q)/\partial Q^2\right]=g_0
\left(1+\beta\mu\lambda^2\omega_b^2\right).
\end{eqnarray}
The higher-order derivative corrections are
$\Theta[W]=\beta D_0g_0\lambda^2
U'(Q)[\partial^3W/\partial Q^3+\beta\partial^2 (U'(Q)W)/\partial Q^2]+
D_0g_0\lambda^2[\partial^4W/\partial Q^4+\beta\partial^3 (U'(Q)W)/\partial Q^3]$.

To estimate the escape rate, we adopt a reduced,
second-order Smoluchowski closure that retains
the drift and diffusion terms in Eq.~(\ref{eqafull})
and neglects the higher-order corrections $\Theta[W]$.
We assess its validity by comparing the predicted rates
with numerical simulations of Eq.~(\ref{eqlan})
(see the Supplemental Material).
Within this reduced description, the averaged evolution
equation becomes $\partial W/\partial t
\simeq(1-\Xi)D_0\{\partial^2W/\partial Q^2
+\beta\partial[U'(Q)W]/\partial Q\}$.
We define the renormalized barrier-crossing diffusivity as $D_{\rm ren}\equiv D_0[1-\Xi]$ and
restrict its application to $0\leq\Xi<1$, ensuring positive diffusivity.
Using the Einstein relation $D_{\rm ren}=(\beta\mu\gamma_{\rm ren})^{-1}$
and $D_0=(\beta\mu\gamma_0)^{-1}$ \cite{kubor1,kubor2}, we obtain the renormalized barrier-crossing friction
\begin{equation}
\label{eqgam}
\gamma_{\rm ren}
=\frac{\gamma_0}{1-\Xi}=\frac{\gamma_0}
{1-g_0(1+\beta\mu\lambda^2\omega_b^2)}.
\end{equation}
For $g_0>0$ and $0<1-\Xi<1$, $\gamma_{\rm ren}/\gamma_0>1$. The homogeneous-friction result is recovered as $g_0\to0$.

This friction enhancement reflects the influence of spatially varying dissipation on activated escape kinetics.
Previous work demonstrated a related effect: hydrodynamic lubrication produces position-dependent
diffusivity and slows colloidal bond dissociation \cite{cness}.
In that model, diffusivity varies deterministically with particle separation.
Here, its variation is represented by spatially correlated
fluctuations associated with the molecular hydration
environment.
Regions of reduced diffusivity can delay passage through
the transition region, without full compensation from regions
of enhanced diffusivity
\cite{santo,grebe}.
Spatial correlations further amplify the fluctuations' influence near the transition state,
as reflected by the $\beta\mu\lambda^2\omega_b^2$ term in Eq.~(\ref{eqgam}).
The hydration-induced diffusivity heterogeneity thus reduces the effective mobility
and enhances the barrier-crossing friction.

\begin{figure}[t]
\centering
\includegraphics[width=0.4\textwidth]{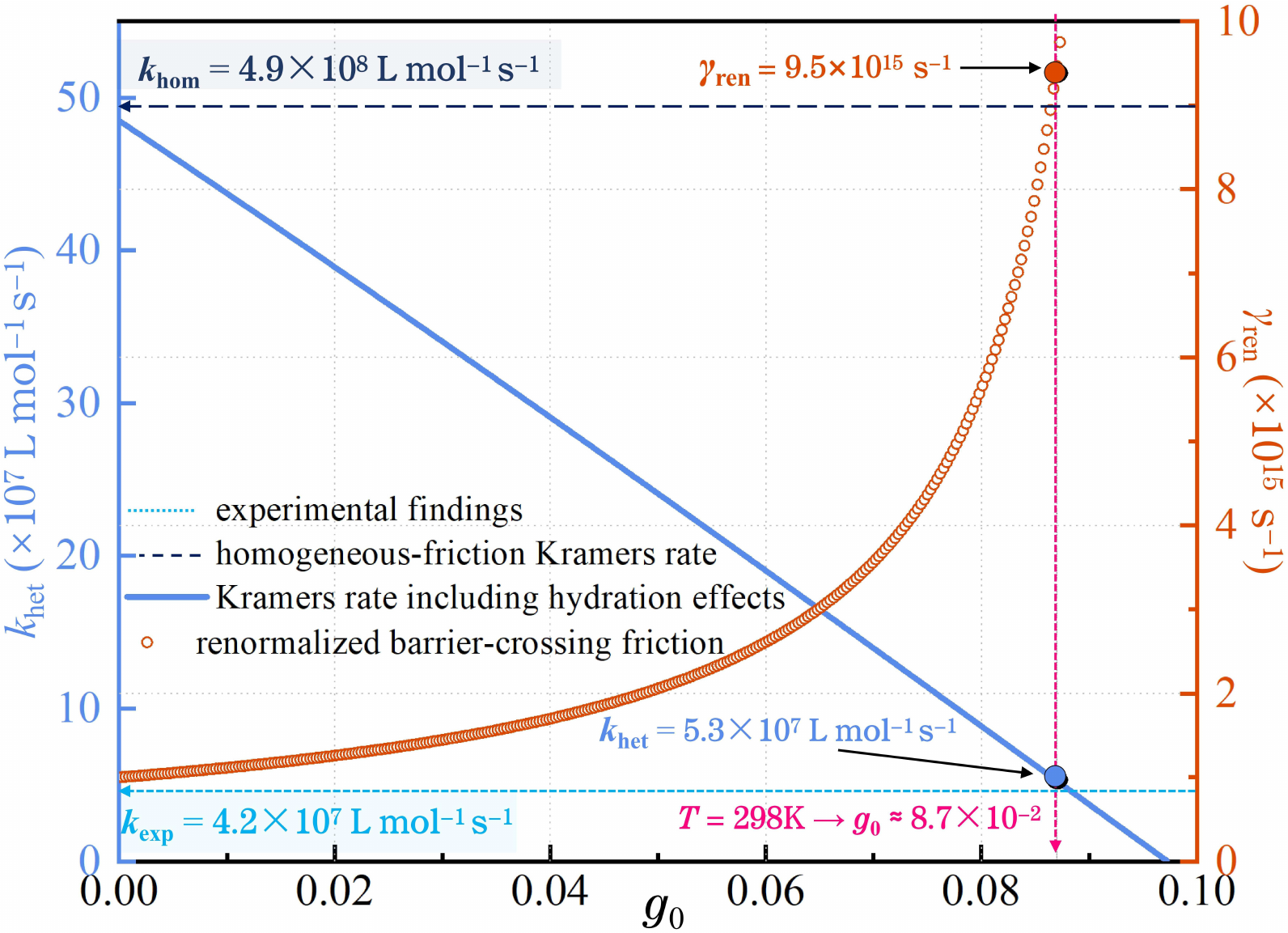}
\caption{\label{recol}
Reduction of the room-temperature Kramers rate discrepancy
by hydration-induced diffusivity heterogeneity.
The horizontal lines denote the homogeneous-friction Kramers rate
constant $k_{\rm hom}$ and the experimental value $k_{\rm exp}$.
With hydration-induced diffusivity heterogeneity included,
the rate constant $k_{\rm het}$ (left axis) and the
renormalized friction $\gamma_{\rm ren}$ (right axis)
are calculated using Eqs.~(\ref{eqkh}) and (\ref{eqgam}),
respectively.
The vertical dashed line corresponds to the hydration-based
estimates $g_0=8.7\times10^{-2}$ and
$\lambda=0.264~\text{\AA}$, obtained from the \ce{^.OH}
hydration environment [Fig.~\ref{hydra}], for which the
discrepancy is substantially reduced.
Other parameters are
$\gamma_0=1.0\times10^{15}\,\mathrm{s^{-1}}$,
$\omega_b=1.72\times10^{14}\,\mathrm{s^{-1}}$, and
$k_{\rm TST}=2.9\times10^9\,
\mathrm{L\,mol^{-1}\,s^{-1}}$. }
\end{figure}

{\it Rate constants and comparison with experiments}.---For reactions in homogeneous
solution, the conventional Kramers
rate constant is $k_{\rm hom}=\kappa k_{\rm TST}$, where $\kappa=\sqrt{1+\gamma_0^2/(4\omega_b^2)}
-\gamma_0/(2\omega_b)$ is the transmission coefficient, and
$k_{\rm TST}$ is the transition-state-theory rate \cite{sicar,polla}.
Hydration-induced diffusivity heterogeneity renormalizes the
barrier-crossing friction from $\gamma_0$ to $\gamma_{\rm ren}$, yielding the rate constant
\begin{equation}
\label{eqkh}
k_{\rm het}=\left[
\sqrt{1+\frac{\gamma_{\rm ren}^2}{4\omega_b^2}}
-\frac{\gamma_{\rm ren}}{2\omega_b}
\right]k_{\rm TST}.
\end{equation}
This substitution provides an effective rate estimate based on the reduced
Smoluchowski description developed above.
In the homogeneous limit $g_0\rightarrow0$,
$\gamma_{\rm ren}\rightarrow\gamma_0$ and
$k_{\rm het}\rightarrow k_{\rm hom}$.
We apply this description to the aqueous
\ce{H2 + ^.OH\rightarrow H^. + H2O} reaction.

(i) Room-temperature comparison.
At $T=298$ K, the experimentally measured value is
$k_{\rm exp}\simeq4.2\times10^7\,
\mathrm{L\,mol^{-1}\,s^{-1}}$ \cite{buxto}.
Reactant and transition-state properties were obtained from unrestricted
$\omega$B97X-D/6-31++G(d,p) calculations, with water described by the IEFPCM
continuum-solvation model at 1 bar.
The unstable transition-state mode has a wavenumber magnitude of
$912~{\rm cm^{-1}}$ and a reduced mass of $1.1155$ amu, corresponding to
$\omega_b\simeq1.72\times10^{14}\ {\rm s^{-1}}$ and
$\mu\simeq1.85\times10^{-27}\ {\rm kg}$.
The corresponding transition-state-theory rate at $298$ K is
$k_{\rm TST}\simeq2.9\times10^{9}\,
{\rm L\,mol^{-1}\,s^{-1}}$ \cite{ctran}.
The barrier height corresponds to $\beta E_b\simeq5.1$.
At this finite barrier height, quantitative deviations
from Kramers theory may occur \cite{azzco}.
For the adopted potential,
the homogeneous-friction Kramers prediction agrees
with direct Langevin simulations to within approximately $2\%$
(Fig.~S1 of the Supplemental Material). This supports its use
as an approximate reference for the adopted model at $298~\mathrm{K}$.
The homogeneous friction coefficient is estimated
as $\gamma_0=6\pi a\eta/\mu\simeq1.0\times10^{15}\ {\rm s^{-1}}$, using $a\simeq1.10~\text{\AA}$ for the
hydrogen-dominated transition-state motion and
$\eta\simeq8.91\times10^{-4}\ {\rm Pa\,s}$ for water at $298$ K.
These parameters give
$k_{\rm hom}\simeq4.9\times10^8\,
\mathrm{L\,mol^{-1}\,s^{-1}}$,
which exceeds the experimental value by more than an order of
magnitude (see Fig.~\ref{recol}).

We next incorporate the diffusivity heterogeneity associated
with the local hydration environment of \ce{^.OH}.
Previous studies report $N_{\rm hyd}\approx13.6$ and
$R_{\rm hyd}\approx3.59~\text{\AA}$ at $298$ K \cite{couto,kazmi}.
These values give
$g_0\sim\Omega(N_{\rm hyd})\approx8.7\times10^{-2}$
[Fig.~\ref{hydra}(b)] and
$\lambda\sim R_{\rm hyd}/N_{\rm hyd}\approx0.264~\text{\AA}$.
Equation~(\ref{eqgam}) then gives
$\gamma_{\rm ren}\simeq9.5\times10^{15}\,\mathrm{s^{-1}}$, yielding
\begin{equation}
\label{eqkhyd}
k_{\rm het}
\simeq
5.3\times10^{7}\,
{\rm L\,mol^{-1}\,s^{-1}}.
\end{equation}
The predicted rate is approximately $26\%$ higher than the experimental
value [Fig.~\ref{recol}], whereas the homogeneous-friction prediction is approximately $11.7$
times the experimental rate.
Thus, within this model, hydration-induced friction
enhancement substantially reduces the room-temperature
discrepancy between the homogeneous-friction prediction
and experiment.

\begin{figure}[t]
\centering
\includegraphics[width=0.42\textwidth]{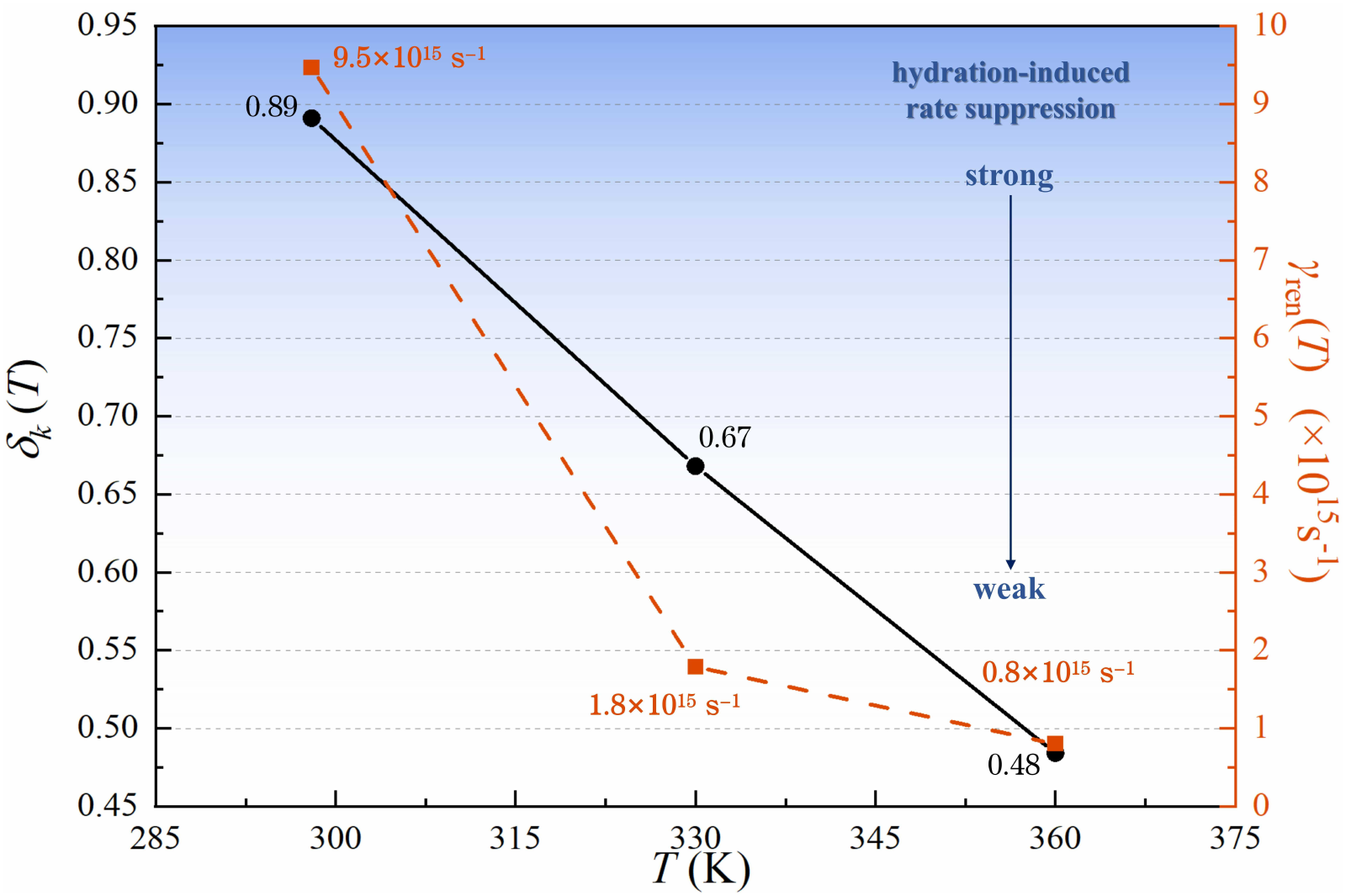}
\caption{\label{tempt}
Temperature dependence of the hydration-induced rate correction.
The fractional rate suppression $\delta_k(T)$ (left axis) decreases
as $T$ increases, accompanied by a decrease in the renormalized friction $\gamma_{\rm ren}(T)$ (right axis) from Eq.~(\ref{eqgamt}).
The hydration-induced effect on the rate constant weakens
progressively as $T$ increases.
For $T=298$, $330$, and $360$ K, respectively,
$\gamma_0\approx(1.0,\,0.55,\,0.36)\times10^{15}\ {\rm s^{-1}}$ and
$g_0\approx(8.7,\,7.4,\,6.3)\times10^{-2}$.
All other parameters are the same as in Fig.~\ref{recol}.}
\end{figure}

(ii) Temperature dependence of the rate suppression.
We next examine the predicted temperature dependence of the
hydration-induced rate suppression over $298$--$360$ K,
a range chosen to keep the aqueous solvent in the liquid
phase at $1$ bar.
Simulations suggest that the mean \ce{^.OH} hydration-cage radius remains
approximately $3.6~\text{\AA}$ over this temperature range \cite{kazmi}.
In this analysis, we hold $N_{\rm hyd}$ and $R_{\rm hyd}$,
and hence $\lambda$, fixed while retaining the temperature
dependence of the local mobility fluctuations.
The parameters $a$, $\mu$, and $\omega_b$ are also held fixed.
The fluctuation analysis in
Eqs.~(\ref{eqome})--(\ref{eqg0}) gives
$g_0(T)\simeq\Omega(N_{\rm hyd},T)$.
As shown in Fig.~\ref{hydra}(b), the estimated fluctuation strength
decreases from $g_0\approx8.7\times10^{-2}$ at $298$ K to
$7.4\times10^{-2}$ at $330$ K, and
$6.3\times10^{-2}$ at $360$ K.
The homogeneous friction varies with the dynamic viscosity
of water according to $\gamma_0(T)=6\pi a\eta(T)/\mu$.
The temperature-dependent dynamic viscosity of water, $\eta(T)$, was evaluated
using the IAPWS correlation for liquid water \cite{hurbj}.
The resulting values are $8.91\times10^{-4}$, $4.89\times10^{-4}$, and
$3.26\times10^{-4}~\mathrm{Pa\,s}$ at $298$, $330$, and $360~\mathrm{K}$, respectively.
The renormalized friction is therefore
\begin{equation}
\label{eqgamt}
\gamma_{\rm ren}(T)=\gamma_0(T)/[1-\Xi(T)],
\end{equation}
where $\Xi(T)=g_0(T)[1+\lambda^2\mu\omega_b^2/(k_BT)]$.
Both the decrease in $g_0(T)$ and the explicit $1/T$ factor
reduce $\Xi(T)$ as temperature increases, thereby weakening the relative
friction enhancement
$\gamma_{\rm ren}(T)/\gamma_0(T)=1/[1-\Xi(T)]$.
The water viscosity sets the homogeneous friction scale and affects the transmission coefficients entering both rates.

To quantify the rate suppression induced by diffusivity heterogeneity
relative to the homogeneous Kramers prediction, we define
$\delta_k(T)\equiv |k_{\rm hom}(T)-k_{\rm het}(T)|/k_{\rm hom}(T)$.
Within the range $0\leq\Xi(T)<1$,
$\gamma_{\rm ren}(T)\geq\gamma_0(T)$ and hence
$k_{\rm het}(T)\leq k_{\rm hom}(T)$.
Because the same $k_{\rm TST}(T)$ enters both rates, it cancels in their ratio, giving
$\delta_k(T)=1-[\sqrt{4\omega_b^2+\gamma_{\rm ren}^2(T)}
-\gamma_{\rm ren}(T)]/[\sqrt{4\omega_b^2+\gamma_0^2(T)}-\gamma_0(T)$].
Here, $\delta_k=0$ corresponds to no heterogeneity-induced
rate suppression, whereas larger $\delta_k$ indicates stronger suppression.
Although $k_{\rm TST}(T)$ cancels, the dependence on $\gamma_0(T)$ remains in the transmission-coefficient ratio.
As shown in Fig.~\ref{tempt}, the predicted fractional
rate suppression decreases from $\delta_k\approx0.89$ at $298$ K
to $0.67$ at $330$ K and $0.48$ at $360$ K.
Thus, the hydration-induced correction remains appreciable over this
temperature range but becomes progressively weaker with increasing temperature.

In summary, we have developed a coarse-grained description of
local diffusivity heterogeneity arising from the \ce{^.OH}
hydration environment for the aqueous reaction
\ce{H2 + ^.OH\rightarrow H^. + H2O}.
We represent this heterogeneity by spatially correlated
diffusivity fluctuations. The fluctuation strength is estimated
from water mobility statistics at the molecular scale set by
the \ce{^.OH} hydration number, and the correlation length
from hydration-cage geometry.
Within this description, diffusivity heterogeneity enhances
the effective barrier-crossing friction and suppresses the
reaction rate, thereby substantially reducing the
room-temperature discrepancy with experiment.
The model further predicts that both the relative friction
enhancement and the rate suppression weaken progressively
with increasing temperature over $298$--$360$ K.
These results provide a possible microscopic link between local hydration dynamics,
barrier-crossing dissipation, and activated reaction kinetics
in aqueous solution.
Further studies of other aqueous reactions could assess the
applicability of these parameter estimates across different
hydration environments, where local solvent heterogeneity
may produce dynamical effects beyond homogeneous-friction
descriptions \cite{dalton,kiefe}.

{\it Acknowledgments.}---This work was supported by the National Natural Science Foundation
of China (Grant No. 12405041), the Guangdong Basic and Applied Basic Research Foundation
(Grant Nos. 2025A1515010709 and 2026A1515010670), and the STU Scientific Research Initiation
Program (Grant No. NTF23014).

{\it Data availability.}---The data that support the findings of this study are
available from the corresponding authors upon reasonable request.


\begin{thebibliography}{99}

\bibitem{dizda} B. Halliwell, A. Adhikary, M. Dingfelder, and M. Dizdaroglu, Hydroxyl radical is a significant player in oxidative DNA damage in vivo, Chem. Soc. Rev. \textbf{50}, 8355 (2021).

\bibitem{suix} X. Sui, J. Wang, Z. Zhao, B. Liu, M. Liu, M. Liu, C. Shi, X. Feng, Y. Fu, D. Shi, \emph{et al.}, Phenolic compounds induce ferroptosis-like death by promoting hydroxyl radical generation in the Fenton reaction, Commun. Biol. \textbf{7}, 199 (2024).

\bibitem{xujk} J. Xu, K. Gu, P. Wang, P. Cheng, H. Che, C. Tang, K. Zhang, and Y. Ao, Piezo-catalytic in-site \ce{H2O2} generation and activation across wide pH range to drive hydroxyl radical-mediated pollutant degradation, Nat. Commun. \textbf{16}, 7908 (2025).

\bibitem{tang} Z. Tang, Y. Liu, M. He, and W. Bu, Chemodynamic therapy: tumour microenvironment-mediated Fenton and Fenton-like reactions, Angew. Chem. Int. Ed. \textbf{58}, 946 (2019).

\bibitem{smith} I. W. M. Smith and F. F. Crim, The chemical kinetics and dynamics of the prototypical reaction:
    \ce{OH + H2 -> H2O + H}, Phys. Chem. Chem. Phys. \textbf{4}, 3543 (2002).

\bibitem{zhang} Z. Zhao, S. Liu, and D. H. Zhang, State-to-state differential cross sections for a four-atom reaction: \ce{H2 + OH\rightarrow H2O + H} in full dimensions, J. Chem. Phys. \textbf{145}, 134301 (2016).

\bibitem{marin} T. W. Marin, C. D. Jonah, and D. M. Bartels, Reaction of \ce{^.OH} radicals with \ce{H2} in sub-critical water, Chem. Phys. Lett. \textbf{371}, 144 (2003).


\bibitem{krame} H. A. Kramers, Brownian motion in a field of force and the diffusion model of chemical reactions, Physica (Amsterdam) \textbf{7}, 284 (1940).

\bibitem{hangg} P. H\"{a}nggi, P. Talkner, and M. Borkovec, Reaction-rate theory: fifty years after Kramers, Rev. Mod. Phys. \textbf{62}, 251 (1990).

\bibitem{garci} P. L. Garc\'{\i}a-M\"{u}ller, F. Borondo, R. Hernandez, and R. M. Benito, Solvent-induced acceleration of the rate of activation of a molecular reaction, Phys. Rev. Lett. \textbf{101}, 178302 (2008).

\bibitem{polla2} E. Pollak, Multidimensional uniform semiclassical instanton thermal rate theory, J. Chem. Phys. \textbf{159}, 224107 (2023).

\bibitem{rouss} M. R. Roussel, {\it Foundations of Chemical Kinetics: A Hands-On Approach} (IOP Publishing, Bristol, 2023), Chap. 12, pp. 12-1--12-25.

\bibitem{buxto} G. V. Buxton, C. L. Greenstock, W. P. Helman, and A. B. Ross, Critical review of rate constants for reactions of hydrated electrons, hydrogen atoms and hydroxyl radicals (\ce{^.OH}/\ce{^.O-}) in aqueous solution, J. Phys. Chem. Ref. Data \textbf{17}, 513 (1988).

\bibitem{siwek} M. Siwek and T. Edgecock, Application of electron beam water radiolysis for sewage sludge treatment--a review, Environ. Sci. Pollut. Res. \textbf{27}, 42424 (2020).

\bibitem{bagch2} S. Mukherjee, S. Mondal, S. Acharya, and B. Bagchi, Tug-of-war between internal and external frictions and viscosity dependence of rate in biological reactions, Phys. Rev. Lett. \textbf{128}, 108101 (2022).

\bibitem{bagch1} S. Mondal and B. Bagchi, From structure and dynamics to biomolecular functions: the ubiquitous role of solvent in biology, Curr. Opin. Struc. Biol. \textbf{77}, 102462 (2022).

\bibitem{brana} B. Rana and J. M. Herbert, Hidden Hemibonding in the Aqueous Hydroxyl Radical, J. Phys. Chem. Lett. \textbf{12}, 8053 (2021).

\bibitem{kjell} L. Kjellsson, K. D. Nanda, J.-E. Rubensson, G. Doumy, S. H. Southworth, P. J. Ho, A. M. March, A. Al Haddad, Y. Kumagai, M.-F. Tu \emph{et al.}, Resonant inelastic X-ray scattering reveals hidden local transitions of the aqueous OH radical, Phys. Rev. Lett. \textbf{124}, 236001 (2020).


\bibitem{codor} E. Codorniu-Hern\'{a}ndez and P. G. Kusalik, Mobility mechanism of hydroxyl radicals in aqueous solution via hydrogen transfer, J. Am. Chem. Soc. \textbf{134}, 532 (2012).

\bibitem{wojci}D. Swiatla-Wojcik and J. Szala-Bilnik, High temperature aqueous solvent effect on translational and hydrogen bond dynamics of the hydroxyl radical-MD simulation study, J. Supercrit. Fluids \textbf{145}, 103 (2019).

\bibitem{hadiz} M. H. Hadizadeh and F. Xu, First results from the interaction between hydroxyl radicals and condensed water droplets: a spin-polarized DFT-MD study, J. Mol. Liq. \textbf{393}, 123557 (2024).

\bibitem{azacc} A. Zaccone, Note: slowing-down of diffusion-controlled reactions in dense liquid matter, J. Chem. Phys. \textbf{138}, 186101 (2013).

\bibitem{addad1} G. R. Haynes, G. A. Voth, and E. Pollak, A theory for the thermally activated rate constant in systems with spatially dependent friction, Chem. Phys. Lett. \textbf{207}, 4 (1993).

\bibitem{addad2} G. R. Haynes, G. A. Voth, and E. Pollak, A theory for the activated barrier crossing rate constant in systems influenced by space and time dependent friction, J. Chem. Phys. \textbf{101}, 7811 (1994).

\bibitem{pgbha} E. Pollak, H. Grabert, and P. H\"{a}nggi, Theory of activated rate processes for arbitrary frequency dependent friction: Solution of the turnover problem, J. Chem. Phys. \textbf{91}, 4073 (1989).

\bibitem{reacc} E. Pollak, A personal perspective of the present status and future challenges facing thermal reaction rate theory, J. Chem. Phys. \textbf{160}, 150902 (2024).

\bibitem{pabis} A. Pabis, J. Szala-Bilnik, and D. Swiatla-Wojcik, Molecular dynamics study of the hydration of the hydroxyl radical at body temperature, Phys. Chem. Chem. Phys. \textbf{13}, 9458 (2011).

\bibitem{grote} R. F. Grote and J. T. Hynes, The stable states picture of chemical reactions. II. rate constants for condensed and gas phase reaction models, J. Chem. Phys. \textbf{73}, 2715 (1980).

\bibitem{eploa} E. Pollak, Theory of activated rate processes: a new derivation of Kramers' expression, J. Chem. Phys. \textbf{85}, 865 (1986).

\bibitem{wolfs1} S. Wolf, B. Lickert, S. Bray, and G. Stock, Multisecond ligand dissociation dynamics from atomistic simulations, Nat. Commun. \textbf{11}, 2918 (2020).

\bibitem{wolfs2} S. Milster, J. Dzubiella, G. Stock, and S. Wolf, Nonequilibrium friction and free energy estimates for kinetic coarse-graining---driven particles in responsive media, J. Chem. Phys. \textbf{162}, 154113 (2025).

\bibitem{mgli1} M. G. Li, M. Hu, L. M. Fan, J. D. Bao, and P. C. Li, Quantifying the energy landscape in weakly and strongly disordered frictional media, J. Chem. Phys. \textbf{160}, 024903 (2024).

\bibitem{azacco} A. Zaccone, N. Dorsaz, F. Piazza, C. De Michele, M. Morbidelli, and G. Foffi, Crowding, intermolecular interactions, and shear flow effects in the diffusion model of chemical reactions, J. Phys. Chem. B \textbf{115}, 7383 (2011).

\bibitem{chers} A. G. Cherstvy, A. V. Chechkin, and R. Metzler, Anomalous diffusion and
ergodicity breaking in heterogeneous diffusion processes, New J. Phys. \textbf{15}, 083039 (2013).
\bibitem{kubor} R. Kubo, M. Toda, and N. Hashitsume, \emph{Statistical Physics II, Nonequilibrium Statistical Mechanics} (Springer, Berlin, 1985).

\bibitem{bruni} F. N. Br\"{u}nig, R. R. Netz, and J. Kappler, Barrier-crossing times for different non-Markovian friction in well and barrier: a numerical study, Phys. Rev. E \textbf{106}, 044133 (2022).

\bibitem{eolla} E. Pollak and S. Miret-Art\'{e}s, Recent developments in Kramers' theory of reaction rates, ChemPhysChem \textbf{24}, e202300272 (2023).
\bibitem{sokol} A. Pacheco-Pozo and I. M. Sokolov, Random walks in correlated diffusivity landscapes, Eur. Phys. J. B \textbf{96}, 152 (2023).

\bibitem{scior} J. M. Montes de Oca, S. R. Accordino, G. A. Appignanesi, P. H. Handle, and F. Sciortino, Size dependence of dynamic fluctuations in liquid and supercooled water, J. Chem. Phys. \textbf{150}, 144505 (2019).

\bibitem{nikka} N. Kikkawa, R. Jinnouchi, T. Nagai, S. Okazaki, and M. Kimura, Local diffusion analysis using square displacement averaged in subspace, J. Chem. Phys. \textbf{164}, 114105 (2026).

\bibitem{madiz} M. H. Hadizadeh, L. Yang, G. Fang, Z. Qiu, and Z. Li, The mobility and solvation structure of a hydroxyl radical in a water nanodroplet: a Born-Oppenheimer molecular dynamics study, Phys. Chem. Chem. Phys. \textbf{23}, 14628 (2021).

\bibitem{riske} H. Risken, \emph{The Fokker-Planck Equation: Methods of Solution and Applications}, 2nd ed. (Springer, Berlin, 1996).

\bibitem{rober} R. B. Best and G. Hummer, Diffusive model of protein folding dynamics with Kramers turnover in rate, Phys. Rev. Lett. \textbf{96}, 228104 (2006).

\bibitem{cness} C. Ness and A. Zaccone, Effect of hydrodynamic interactions on the lifetime of colloidal bonds, Ind. Eng. Chem. Res. \textbf{56}, 3726 (2017).

\bibitem{klayt} V. I. Klyatskin, \emph{Statistical Description of Dynamical Systems with Fluctuating Parameters} (Nauka, Moscow, 1975) (in Russian).

\bibitem{santo} M. A. F. dos Santos, L. Menon, and C. Anteneodo, Efficiency of random search with space-dependent diffusivity, Phys. Rev. E \textbf{106}, 044113 (2022).

\bibitem{grebe} Y. Lanoisel\'{e}e, N. Moutal, and D. S. Grebenkov, Diffusion-limited reactions in dynamic heterogeneous media, Nat. Commun. \textbf{9}, 4398 (2018).
\bibitem{kubor1} R. Kubo, The fluctuation-dissipation theorem, Rep. Prog. Phys. \textbf{29}, 255 (1966).

\bibitem{kubor2} A. Baer, S. E. Wawra, K. Bielmeier, M. J. Uttinger, D. M. Smith, W. Peukert, J. Walter, and A.-S. Smith, The Stokes-Einstein-Sutherland equation at the nanoscale revisited, Small \textbf{20}, 2304670 (2024).

\bibitem{sicar} F. Sicard, Computing transition rates for rare events: when Kramers theory meets the free-energy landscape, Phys. Rev. E \textbf{98}, 052408 (2018).

\bibitem{polla} E. Pollak, Instanton-based Kramers turnover theory, Phys. Rev. A \textbf{109}, 022242 (2024).

\bibitem{ctran} At 298 K, the transition-state-theory rate constant is calculated as $k_{\text{TST}}=\frac{k_BT}{hc^\circ}\frac{\hat q_T}{\hat q_{\mathrm{H_2}}\hat q_{\mathrm{OH}}}\exp
    \left(-\frac{E_b}{k_BT}\right)$, where $h$ is Planck's constant, and $c^\circ=1\ \mathrm{mol/m^{3}}$ is the standard-state concentration. The calculated energies of \ce{H2}, \ce{^.OH}, and the transition state are $-1.175435$, $-75.717467$, and $-76.888087$ hartree, respectively, giving an activation barrier of $E_b=4.815\times10^{-3}$ hartree =$2.10\times10^{-20}$J. The corresponding partition functions are $\hat q_{\mathrm{H_2}}=4.298$, $\hat q_{\mathrm{OH}}=6.727\times10^3$, and $\hat q_T=2.236$. Substituting these values yields $k_{\rm TST}=2.9\times10^9\ \mathrm{L\ mol^{-1}\ s^{-1}}$.

\bibitem{azzco} M. Abkenar, T. H. Gray, and A. Zaccone, Dissociation rates from single-molecule pulling experiments under large thermal fluctuations or large applied force, Phys. Rev. E \textbf{95}, 042413 (2017).

\bibitem{couto} P. Cabral do Couto, R. C. Guedes, B. J. Costa Cabral, J. A. Martinho Sim\~{o}es, and J. A. M. Sim\~{o}es, The hydration of the OH radical: microsolvation modeling and statistical mechanics simulation, J. Chem. Phys. \textbf{119}, 7344 (2003).

\bibitem{kazmi} L. Kazmierczak, J. Szala-Rearick, and D. Swiatla-Wojcik, 3D characterization of the molecular neighborhood of \ce{^.OH} radical in high temperature water by MD simulation and Voronoi polyhedra, Int. J. Mol. Sci. \textbf{24}, 3294 (2023).

\bibitem{hurbj} J. P\'{a}tek, J. Hrub\'{y}, J. Klomfar, M. Sou\v{c}kov\'{a}, and A. H. Harvey, Reference correlations for thermophysical properties of liquid water at 0.1 MPa, J. Phys. Chem. Ref. Data \textbf{38}, 21 (2009).

\bibitem{dalton} B. A. Dalton, H. Kiefer, and R. R. Netz, The role of memory-dependent friction and solvent viscosity in isomerization kinetics in viscogenic media, Nat. Commun. \textbf{15}, 3761 (2024).

\bibitem{kiefe} H. Kiefer, D. Vitali, B. A. Dalton, L. Scalfi, and R. R. Netz, Effect of frequency-dependent shear and volume viscosities on molecular friction in liquids, Phys. Rev. E \textbf{111}, 015104 (2025).

\end{thebibliography}
\end{document}